\documentclass[amssymb,useAMS,prd,aps,amsmath,twocolumn,superscriptaddress,,nofootinbib]{revtex4}
\usepackage{pslatex}
\usepackage{graphicx}
\usepackage{psfrag}

\def\refe@jnl#1{{#1}}
\def\aj{\refe@jnl{Astron.~J.}}
\def\araa{\refe@jnl{Annu.~Rev.~Astron.~Astrophys.}}
\def\apj{\refe@jnl{Astrophys.~J.}}
\def\apjl{\refe@jnl{Astrophys.~J.~Lett.}}
\def\aap{\refe@jnl{Astron.~Astrophys.}}
\def\mnras{\refe@jnl{Mon.~Not.~R.~Astron.~Soc.}}
\def\prd{\refe@jnl{Phys.~Rev.~D}}
\def\fcp{\refe@jnl{Fund.~Cos.~Phys.}}
\def\physrep{\refe@jnl{Phys.~Rep.}}
\def\physlett{\refe@jnl{Phys.~Lett.}}

\def\invisible#1{  }

\def\dm{{\rm dm}}

\def\mdm{{m_\dm}}

\begin{document}

\title{A new test of the light dark matter hypothesis}

\author{C\'eline B\oe hm}

\affiliation{LAPTH, UMR 5108, 9 chemin de Bellevue - BP 110,
  74941 Annecy-Le-Vieux, France.}

\author{Joseph Silk}
\affiliation{Astrophysics department, DWB Keble road, OX1 3RH
Oxford, UK}

\date{today}

\begin{abstract}
Detection of a surprisingly high flux of positron annihilation
radiation from the inner galaxy has motivated the proposal that dark
matter is made of weakly interacting light particles (possibly as
light as the electron). This scenario is extremely hard to test in
current high energy physics experiments. Here, however, we
demonstrate
  that the current value of the electron anomalous magnetic moment
already has the required precision to unambiguously test the light
dark  matter hypothesis. If confirmed, the implications for
astrophysics are far-reaching.
\end{abstract}
\maketitle

\section{Introduction}
Many cosmological observations have shown that visible matter only
represents 20$\%$ of the mass of the Universe. The remaining 80$\%$
is of unknown origin. For several decades, particle physicists have
speculated that this ``dark'' matter was made of very heavy
particles (from one to a thousand  proton masses). Recently,
however, the confirmation that low energy positrons are being
emitted in the centre of the galaxy \cite{Jean:2003ci} motivated the
proposal that light particle annihilations could account for the
extended 511 keV line emission \cite{Boehm:2003bt} detected since
the 1970s. Dark matter would then be made of light scalars
annihilating via the exchange of both a new gauge boson $Z'$ and a
heavy fermion $F_e$. The existence of  the $F_e$ particle would
explain the morphology of the 511 keV line while the $Z'$ would
explain why dark matter represents $80\%$ of the matter content of
our universe. Both particles are very hard to test in current
particle physics experiments. Therefore the question arises as to
whether one can ever unambiguously test this model in the
laboratory.

A few months ago, a group proposed an electron-proton scattering
experiment to detect light new gauge bosons
\cite{Heinemeyer:2007sq}. Although such an experiment certainly
deserves to be done, it may not unambiguously determine the source
of antimatter in the Milky Way. Indeed, not only is the $Z'$ not
directly related to the 511 keV line but there is also an
alternative scenario in which the $Z'$ is replaced by a neutral
fermion $N$ \cite{Boehm:2006mi}. Hence, the only way to really
answer the question of the light particle annihilation origin of low
energy positrons in our galaxy is to directly test the existence of
$F_e$ particles. Here we present such a test.


\section{Electron anomalous magnetic moment}

With the latest observations by the SPI spectrometer on the
INTEGRAL satellite, it became possible to establish that the 511 keV
line (already observed by five experiments and now detected at
16$\sigma$) originates from electron-positron pair annihilations at
rest. The rate at which these positrons are being emitted and their
spatial distribution are extremely hard to explain with conventional
and/or new (as yet unknown) astrophysical sources. On the other hand,
both the energy and spatial distribution is suggestive of  dark matter
characteristics. Hence, one hypothesis is that SPI has
 discovered,  for the first time, evidence that dark matter is made of nonbaryonic particles.

To explain the observed morphology of the 511 keV line, dark matter
 must be light (a thousand to a hundred times lighter than a proton, i.e. with a mass comparable to the electron mass)
 and must annihilate into electron-positron pairs. The model which was first
proposed relies on $Z'$ and $F_e$ exchanges. In the primordial
universe, the $Z'$ reduces the dark matter abundance to the observed
value while the $F_e$ provides a subdominant source of positrons. In
the Milky Way, the importance of these two particles is inverted
(due to the velocity-dependence of the $Z'$ exchange cross-section)
and the $F_e$ becomes the dominant source of electron-positron pairs.

Since both $Z'$ and $F_e$ interact with electrons, they are expected
to modify the electron characteristics to some extrent. For example,
the Land\'e factor $g_e$ which is  equal to $g_e = 2 \ (1+a_e)$,
with $a_e$ the so-called anomalous magnetic moment of the electron
due to QED, QCD, weak interactions, could receive two additional
contributions $\delta g_e^{Z',F} = 2 a_e^{Z',F}$ with:
\\
\\
\begin{tabular}{|c|c|c|}
  \hline
    & $F_e$ & $Z'$ \\
  \hline
  \hline
  \ $a_e$ \ & $\frac{c_l c_r \ m_e}{16 \, \pi^2 \, m_{F_e}}$
 &$\frac{z_e^2}{12 \, \pi^2} \frac{m_e^2}{m_{Z'}^2}$  \ \\
 \ = \ &$ \ 5. \, 10^{-12} \ \sqrt{f} \
\left(\frac{\mdm}{\rm{MeV}}\right)$ \ & $ \ 10^{-11} \
\left(\frac{z_e}{7 \, 10^{-5}}\right)^2 \,
\left(\frac{m_{Z'}}{\rm{MeV}}\right)^{-2} \,
$ \ \\
 \hline
\end{tabular}
\\
\\
where we use the expressions and values of the $F_e$
\cite{Boehm:2004gt,Ascasibar:2005rw} and $Z'$ exchange
\cite{Boehm:2003hm} annihilation cross-sections. The parameter  $f$
reflects the type of dark matter halo profile that fits the data. As
shown in Ref.~\cite{Ascasibar:2005rw}, it is equal to unity in the
case of a Navarro-Frenk-White profile (much flatter or cuspier
profiles were ruled out by this analysis).

The remarkable point is that $a_e^F$ only depends on the morphology
of the 511 keV line and is therefore constrained to lie within a
narrow range of values. This enables us to make firm predictions.
So far, the $Z'$ contribution, which seems closer to the
experimental value of $(g_e-2)$,  was always regarded as more
important. Here, however, we point out that the $F_e$ particles
 give the dominant contribution to the electron anomalous magnetic moment. This opens up
new possibilities to test the light dark matter scenario in
laboratory experiments.

\begin{figure}
  \includegraphics[width=9.5cm]{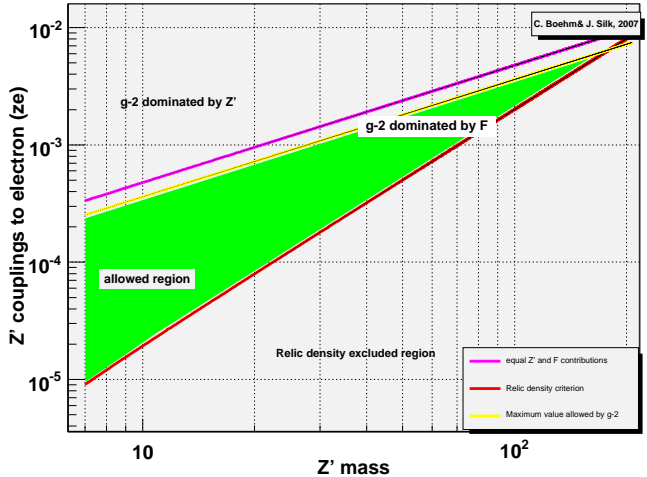}\\
  \caption{Here we plotted the $Z'$ coupling to electrons versus the $Z'$ mass for $\mdm= 1$ MeV
  and $(a_e^F + a_e^{Z'})
\leq \delta a_{Rb06}$ (we do not plot the error bars).
  The allowed parameter space is represented in green. The upper region (above the magenta
  line) represents $z_e$ values for which the
$Z'$ contribution to the electron anomalous magnetic moment is
dominant. As one can see, $a_e^{F}$  always dominates over
$a_e^{Z'}$ in the allowed region. This region shrinks towards the
red line for increasing values of $\mdm$.}\label{fig1}
\end{figure}

The term $a_e^{Z'}$ depends on two parameters: the mass $m_{Z'}$ and
the coupling to electron $z_e$. The latter varies from a minimal
value $z_e^{min, RD}$, fixed by the relic density condition, to a
maximal value $z_e^{max, g_e-2}$, fixed by the condition $a_e^{Z'} =
a_e^{exp}$ (with $a_e^{exp}$ the measured value of the the anomalous
magnetic moment)\cite{Boehm:2003hm}. The $Z'$ contribution to the
electron anomalous magnetic moment  dominates over the $F_e$
contribution  when
$$z_e > z_e^{eq} = 4.7 \ 10^{-5} \ \sqrt{f} \
\left(\frac{m_{Z'}}{\rm{MeV}}\right) \ \sqrt{\left(\frac{\mdm}{\rm{1
MeV}}\right)}.$$ However, as we demonstrate in Fig.~\ref{fig1} for
$\mdm=1$ MeV, and varying $m_{Z'}$ between 7 to 200 MeV, we find
that $z_e^{eq}$ is always outside the allowed range, {\it i.e.}
$a_e^F$ \textbf{is} the main extra contribution to the electron
$g_e-2$ value. One can therefore unambiguously test the existence of
$F_e$ particles and elucidate the origin of the low energy positrons
in the Milky Way by using $g_e-2$ measurements.

 The latest
measurement of the electron anomalous magnetic moment by the Harvard group  in 2006 \cite{Gabrielse:2006gg},  using a One-Electron
Quantum Cyclotron,  gives $a_e^{exp, 06}= g/2-1 = 0.001 \, 159 \, 652
\, 180 \, 85 (76)$, with a $7.6 \ 10^{-13}$ uncertainty. We estimate
the $F_e$ contribution to lie between
$$a_e^{F} \ \in \ [5 \ \sqrt{f} \ 10^{-12}\ , \ 1.5 \ \sqrt{f} \ 10^{-11}],$$
using dark matter mass range derived in \cite{Beacom:2005qv}. This
is comparable to the level of precision that has been reached
experimentally. This experiment is therefore sensitive enough to
prove $F_e$ interactions.

Let us now compare our estimate with the discrepancy between the
 $a_e^{exp}$ experimental value  and the Standard Model prediction
(without light dark matter). The theoretical estimate $a_e^{th}$
depends on one parameter: the value of the fine structure constant
$\alpha$. Unfortunately, this is not known  with the same level of
precision as the electron anomalous magnetic moment but the latest
(most precise) measurements of $\alpha$ by the Caesium and Rydberg
constant experiments in 2006 \cite{Clade,gerginov} nevertheless lead
to the following differences \cite{Gabrielse:2006gg}:
$$\delta a_{CS06} = a^{th}_e(\alpha_{Cs06}) - a^{exp, 06}_{e} = - 7.9 \ (9.3)
\ 10^{-12}$$ and
$$\delta a_{Rb06}  = a^{th}_e(\alpha_{Rb06}) - a^{exp, 06}_{e} = 1.9 \ (7.7) \
10^{-12}$$ respectively. At present, this comparison only indicates
that there is room for light dark matter. However, as soon as
measurements of $\alpha$ will converge and the error bars will
decrease, one should be able to determine whether or not scalar
particles coupled to heavy fermions $F_e$ can be responsible for the
emission of the 511 keV line in our galaxy.

Note that the relation between $a_e^{th}$ and the fine structure
constant includes calculations of QED, WEAK and QCD contributions
(QED to the tenth-order \cite{kinoshita}). The next order
calculations may change the present discrepancies and even be
sufficient to solve the dark matter problem.

\section{Conclusion}
Using the 511 keV line properties, we have shown that the $F_e$
contribution to the electron anomalous magnetic moment was dominant
and within the reach of the present anomalous magnetic moment
experiment. It is remarkable that  precise measurement of the
fine structure constant offers the best way to test the light dark
matter scenario. Although such a test requires the experimental and
computational expertise of particle physicists, the result should
 have a great impact on the astrophysics/astroparticle
community by answering whether or not dark matter is light and
 has actually been discovered by SPI. A negative answer
would indicate that the origin of the low energy positrons in our
galaxy is of (hitherto unknown) astrophysical origin. This would in
turn motivate high sensitivity point source searches as well as
improved diffuse X-ray/$\gamma$-ray background observations.

\bibliography{gmdb}

\begin{thebibliography}{12}
\expandafter\ifx\csname natexlab\endcsname\relax\def\natexlab#1{#1}\fi
\expandafter\ifx\csname bibnamefont\endcsname\relax
  \def\bibnamefont#1{#1}\fi
\expandafter\ifx\csname bibfnamefont\endcsname\relax
  \def\bibfnamefont#1{#1}\fi
\expandafter\ifx\csname citenamefont\endcsname\relax
  \def\citenamefont#1{#1}\fi
\expandafter\ifx\csname url\endcsname\relax
  \def\url#1{\texttt{#1}}\fi
\expandafter\ifx\csname urlprefix\endcsname\relax\def\urlprefix{URL }\fi
\providecommand{\bibinfo}[2]{#2}
\providecommand{\eprint}[2][]{\url{#2}}

\bibitem[{\citenamefont{Jean et~al.}(2003)}]{Jean:2003ci}
\bibinfo{author}{\bibfnamefont{P.}~\bibnamefont{Jean}} \bibnamefont{et~al.},
  \bibinfo{journal}{Astron. Astrophys.} \textbf{\bibinfo{volume}{407}},
  \bibinfo{pages}{L55} (\bibinfo{year}{2003}), \eprint{astro-ph/0309484}.

\bibitem[{\citenamefont{Boehm et~al.}(2004)\citenamefont{Boehm, Hooper, Silk,
  Casse, and Paul}}]{Boehm:2003bt}
\bibinfo{author}{\bibfnamefont{C.}~\bibnamefont{Boehm}},
  \bibinfo{author}{\bibfnamefont{D.}~\bibnamefont{Hooper}},
  \bibinfo{author}{\bibfnamefont{J.}~\bibnamefont{Silk}},
  \bibinfo{author}{\bibfnamefont{M.}~\bibnamefont{Casse}}, \bibnamefont{and}
  \bibinfo{author}{\bibfnamefont{J.}~\bibnamefont{Paul}},
  \bibinfo{journal}{Phys. Rev. Lett.} \textbf{\bibinfo{volume}{92}},
  \bibinfo{pages}{101301} (\bibinfo{year}{2004}), \eprint{astro-ph/0309686}.

\bibitem[{\citenamefont{Heinemeyer et~al.}(2007)\citenamefont{Heinemeyer, Kahn,
  Schmitt, and Velasco}}]{Heinemeyer:2007sq}
\bibinfo{author}{\bibfnamefont{S.}~\bibnamefont{Heinemeyer}},
  \bibinfo{author}{\bibfnamefont{Y.}~\bibnamefont{Kahn}},
  \bibinfo{author}{\bibfnamefont{M.}~\bibnamefont{Schmitt}}, \bibnamefont{and}
  \bibinfo{author}{\bibfnamefont{M.}~\bibnamefont{Velasco}}
  (\bibinfo{year}{2007}), \eprint{arXiv:0705.4056 [hep-ex]}.

\bibitem[{\citenamefont{Boehm et~al.}(2006)\citenamefont{Boehm, Farzan, Hambye,
  Palomares-Ruiz, and Pascoli}}]{Boehm:2006mi}
\bibinfo{author}{\bibfnamefont{C.}~\bibnamefont{Boehm}},
  \bibinfo{author}{\bibfnamefont{Y.}~\bibnamefont{Farzan}},
  \bibinfo{author}{\bibfnamefont{T.}~\bibnamefont{Hambye}},
  \bibinfo{author}{\bibfnamefont{S.}~\bibnamefont{Palomares-Ruiz}},
  \bibnamefont{and} \bibinfo{author}{\bibfnamefont{S.}~\bibnamefont{Pascoli}}
  (\bibinfo{year}{2006}), \eprint{hep-ph/0612228}.

\bibitem[{\citenamefont{Boehm and Ascasibar}(2004)}]{Boehm:2004gt}
\bibinfo{author}{\bibfnamefont{C.}~\bibnamefont{Boehm}} \bibnamefont{and}
  \bibinfo{author}{\bibfnamefont{Y.}~\bibnamefont{Ascasibar}},
  \bibinfo{journal}{Phys. Rev.} \textbf{\bibinfo{volume}{D70}},
  \bibinfo{pages}{115013} (\bibinfo{year}{2004}), \eprint{hep-ph/0408213}.

\bibitem[{\citenamefont{Ascasibar et~al.}(2006)\citenamefont{Ascasibar, Jean,
  Boehm, and Knoedlseder}}]{Ascasibar:2005rw}
\bibinfo{author}{\bibfnamefont{Y.}~\bibnamefont{Ascasibar}},
  \bibinfo{author}{\bibfnamefont{P.}~\bibnamefont{Jean}},
  \bibinfo{author}{\bibfnamefont{C.}~\bibnamefont{Boehm}}, \bibnamefont{and}
  \bibinfo{author}{\bibfnamefont{J.}~\bibnamefont{Knoedlseder}},
  \bibinfo{journal}{Mon. Not. Roy. Astron. Soc.}
  \textbf{\bibinfo{volume}{368}}, \bibinfo{pages}{1695} (\bibinfo{year}{2006}),
  \eprint{astro-ph/0507142}.

\bibitem[{\citenamefont{Boehm and Fayet}(2004)}]{Boehm:2003hm}
\bibinfo{author}{\bibfnamefont{C.}~\bibnamefont{Boehm}} \bibnamefont{and}
  \bibinfo{author}{\bibfnamefont{P.}~\bibnamefont{Fayet}},
  \bibinfo{journal}{Nucl. Phys.} \textbf{\bibinfo{volume}{B683}},
  \bibinfo{pages}{219} (\bibinfo{year}{2004}), \eprint{hep-ph/0305261}.

\bibitem[{\citenamefont{Gabrielse et~al.}(2006)\citenamefont{Gabrielse,
  Hanneke, Kinoshita, Nio, and Odom}}]{Gabrielse:2006gg}
\bibinfo{author}{\bibfnamefont{G.}~\bibnamefont{Gabrielse}},
  \bibinfo{author}{\bibfnamefont{D.}~\bibnamefont{Hanneke}},
  \bibinfo{author}{\bibfnamefont{T.}~\bibnamefont{Kinoshita}},
  \bibinfo{author}{\bibfnamefont{M.}~\bibnamefont{Nio}}, \bibnamefont{and}
  \bibinfo{author}{\bibfnamefont{B.}~\bibnamefont{Odom}},
  \bibinfo{journal}{Phys. Rev. Lett.} \textbf{\bibinfo{volume}{97}},
  \bibinfo{pages}{030802} (\bibinfo{year}{2006}).

\bibitem[{\citenamefont{Beacom and Yuksel}(2006)}]{Beacom:2005qv}
\bibinfo{author}{\bibfnamefont{J.~F.} \bibnamefont{Beacom}} \bibnamefont{and}
  \bibinfo{author}{\bibfnamefont{H.}~\bibnamefont{Yuksel}},
  \bibinfo{journal}{Phys. Rev. Lett.} \textbf{\bibinfo{volume}{97}},
  \bibinfo{pages}{071102} (\bibinfo{year}{2006}), \eprint{astro-ph/0512411}.

\bibitem[{\citenamefont{Clad\'e et~al.}(2006)\citenamefont{Clad\'e,
  de~Mirandes, Cadoret, Guellati-Khélifa, Schwob, Nez, L., and
  Biraben}}]{Clade}
\bibinfo{author}{\bibfnamefont{P.}~\bibnamefont{Clad\'e}},
  \bibinfo{author}{\bibfnamefont{E.}~\bibnamefont{de~Mirandes}},
  \bibinfo{author}{\bibfnamefont{M.}~\bibnamefont{Cadoret}},
  \bibinfo{author}{\bibfnamefont{S.}~\bibnamefont{Guellati-Khélifa}},
  \bibinfo{author}{\bibfnamefont{C.}~\bibnamefont{Schwob}},
  \bibinfo{author}{\bibfnamefont{F.}~\bibnamefont{Nez}},
  \bibinfo{author}{\bibfnamefont{J.}~\bibnamefont{L.}}, \bibnamefont{and}
  \bibinfo{author}{\bibfnamefont{F.}~\bibnamefont{Biraben}},
  \bibinfo{journal}{Phys.Rev.Lett.} \textbf{\bibinfo{volume}{96}},
  \bibinfo{pages}{033001} (\bibinfo{year}{2006}).

\bibitem[{\citenamefont{Gerginov et~al.}(2006)\citenamefont{Gerginov, Calkins,
  Tanner, McFerran, Diddams, Bartels, and Hollberg}}]{gerginov}
\bibinfo{author}{\bibfnamefont{V.}~\bibnamefont{Gerginov}},
  \bibinfo{author}{\bibfnamefont{K.}~\bibnamefont{Calkins}},
  \bibinfo{author}{\bibfnamefont{C.~E.} \bibnamefont{Tanner}},
  \bibinfo{author}{\bibfnamefont{J.~J.} \bibnamefont{McFerran}},
  \bibinfo{author}{\bibfnamefont{S.}~\bibnamefont{Diddams}},
  \bibinfo{author}{\bibfnamefont{A.}~\bibnamefont{Bartels}}, \bibnamefont{and}
  \bibinfo{author}{\bibfnamefont{L.}~\bibnamefont{Hollberg}},
  \bibinfo{journal}{Phys.Rev. A} \textbf{\bibinfo{volume}{73}},
  \bibinfo{pages}{032504} (\bibinfo{year}{2006}).

\bibitem[{\citenamefont{Kinoshita and Nio}(2006)}]{kinoshita}
\bibinfo{author}{\bibfnamefont{T.}~\bibnamefont{Kinoshita}} \bibnamefont{and}
  \bibinfo{author}{\bibfnamefont{M.}~\bibnamefont{Nio}},
  \bibinfo{journal}{Phys. Rev.} \textbf{\bibinfo{volume}{D73}},
  \bibinfo{pages}{053007} (\bibinfo{year}{2006}), \eprint{hep-ph/0512330}.

\end{thebibliography}

\end{document}